\documentclass[twocolumn, superscriptaddress, prb, showpacs]{revtex4-1} 
\newif\ifGALLEYversion\GALLEYversionfalse

\usepackage{ifthen} 
\usepackage{graphicx}
\usepackage{epstopdf} 
\usepackage{amsmath}

\ifthenelse{\boolean{GALLEYversion}}{
    \marginparwidth 2.7in
    \marginparsep 0.5in
    \def\dor#1{\marginpar{\small DOR: #1}}
}{
    \def\dor#1{\relax}
}

\ifthenelse{\boolean{GALLEYversion}}{
    \marginparwidth 2.7in
    \marginparsep 0.5in
    \def\aam#1{\marginpar{\small AAM: #1}}
}{
    \def\aam#1{\relax}
}

\ifthenelse{\boolean{GALLEYversion}}{
    \marginparwidth 2.7in
    \marginparsep 0.5in
    \def\mcp#1{\marginpar{\small MCP: #1}}
}{
    \def\mcp#1{\relax}
}

\begin{document}

\title{Generalized Wannier functions:
a comparison of molecular electric dipole polarizabilities}

\author{David D. O'Regan}
\email{david.oregan@epfl.ch}
\affiliation{Cavendish Laboratory, University of Cambridge, J. J. Thomson Avenue,
Cambridge CB3 0HE, United Kingdom}
\affiliation{	Theory and Simulation of Materials, 
\'{E}cole Polytechnique F\'{e}d\'{e}rale de Lausanne, 
1015 Lausanne, Switzerland}
\author{ Mike C. Payne}
\affiliation{Cavendish Laboratory, University of Cambridge, J. J. Thomson Avenue,
Cambridge CB3 0HE, United Kingdom}
\author{Arash A. Mostofi}
\affiliation{The Thomas Young Centre and the Department of Materials, 
Imperial College London, London SW7 2AZ,
United Kingdom}
\date{\today}

\begin{abstract}
Localized Wannier functions provide an efficient and
intuitive means by which to compute dielectric properties from 
first principles. They are most commonly constructed in a post-processing step,
following total-energy minimization. Nonorthogonal
generalized Wannier functions (NGWFs)~\cite{PhysRevB.66.035119,*onetep1} 
may also be  optimized
\emph{in situ}, 
in the process of solving for the ground-state density. 
We explore the relationship between 
NGWFs and orthonormal,
maximally localized Wannier functions
(MLWFs)~\cite{PhysRevB.56.12847,*PhysRevB.65.035109},
demonstrating that NGWFs may be used to 
compute electric dipole polarizabilities efficiently, with no 
necessity for post-processing optimization,
and with an accuracy comparable to MLWFs.
\end{abstract}


\maketitle

In this Brief Report, we explore the 
equivalence between nonorthogonal generalized
Wannier functions 
(NGWFs)~\cite{PhysRevB.66.035119,*onetep1},
generated using linear-scaling 
Kohn-Sham density functional theory 
(DFT)~\cite{PhysRev.136.B864,*PhysRev.140.A1133},
and their orthonormal counterparts,
particularly maximally localized Wannier functions 
(MLWFs)~\cite{PhysRevB.56.12847, 
*PhysRevB.65.035109},
both recently reviewed in Ref.~\onlinecite{wannierreview}.
We demonstrate the comparable, high accuracy of the
two formalisms for dielectric response,
laying the foundation for large-scale 
calculation of optical properties.

We begin with the single-particle 
density-matrix defined, for a given
set of Bloch orbitals
$ \lvert  \psi_{n \mathbf{k}} \rangle$, 
where $n$ indexes occupied bands, $\mathbf{k}$ is
the crystal
wave-vector and we suppress 
the spin index for notational clarity, by
\begin{equation}
\hat{\rho} = 
 \sum_{n } \int_{1\rm{BZ}} d\mathbf{k} \; 
\lvert \psi_{n \mathbf{k}}  \rangle
 f_{n \mathbf{k} }
\langle \psi_{n \mathbf{k} } \rvert.
\end{equation}
Here, $1\rm{BZ}$ is the first Brillouin zone corresponding
to the periodic unit cell of volume $V_{\rm cell}$.
A reformulation of such Bloch states
suitable for the study of spatially localized properties
was proposed by Wannier~\cite{PhysRev.52.191}, whose 
eponymously named functions are defined, 
for a unit cell at the lattice vector $\mathbf{R}$, by
\begin{equation} \label{Eq:mywannier}
\lvert w_{ n \mathbf{R} }  \rangle = 
\sqrt{ \frac{V_{\rm cell}}{\left( 
 2 \pi \right)^3 } }
 \int_{1\rm{BZ}} d\mathbf{k} \; e^{- i \mathbf{k} \cdot \mathbf{R} } 
\lvert \psi_{n \mathbf{k}}  \rangle.
\end{equation}
The  orthonormality of Bloch orbitals is preserved, 
\begin{equation}
\langle w_{n \mathbf{R} } \rvert 
w_{m \mathbf{R'} } \rangle
= \delta_{n m} \delta_{\mathbf{R} \mathbf{R'}},
\end{equation}
and we may choose the gauge freely,
so that any prior unitary transformation 
among the orbitals, 
$\lvert \tilde{\psi}_{n \mathbf{k}}\rangle =  
\sum_{m} \lvert \psi_{m \mathbf{k}}\rangle 
U_{m n \mathbf{k}} $,
may also give rise to a valid set of
generalized Wannier functions, via Eq.~\ref{Eq:mywannier}.
 Unoccupied states may be
included in the wannierization, while
maintaining the same occupied density, by appropriately transforming 
the occupancy of the orbitals to give 
 $f_{n \mathbf{k}} $, to give 
  $\tilde{f}_{n m \mathbf{k}} = \sum_{p}
U^{ \dagger}_{n p \mathbf{k}} 
 f_{p \mathbf{k}} 
U_{p m \mathbf{k}}  $.
The density-matrix may be readily expressed
in terms of Wannier functions, in the separable form 
proposed in Ref.~\onlinecite{RevModPhys.32.335}, and given by
\begin{subequations}
\begin{align} \label{Eq:myreal}
\hat{\rho} &{}= 
 \sum_{\mathbf{R} \mathbf{R'} } 
\lvert w_{n \mathbf{R} } \rangle
 k_{n m 
\mathbf{R'} - \mathbf{R} }
\langle w_{m \mathbf{R'} } \rvert , \quad \mbox{where} \\
k_{n m 
 \mathbf{R} } &{}= 
 \frac{V_{\rm cell}}{\left( 
 2 \pi \right)^3 } 
 \int_{1\rm{BZ}} d\mathbf{k} \; e^{- i \mathbf{k} \cdot \mathbf{R} } 
\tilde{f}_{n m \mathbf{k}}, \end{align}
\end{subequations}
is commonly known as the \emph{density kernel}.

The extension of this formalism to
nonorthogonal generalized Wannier functions (NGWFs)
is both of practical interest and utility.
Orthonormality and spatial localization are generally
competing requirements~\cite{PhysRevLett.21.13}, hence nonorthogonal orbitals 
may form a more efficient basis in which to expand 
short-ranged operators and, as a result,
they are used extensively in linear-scaling
DFT approaches.
We may express these NGWFs, 
$\lvert \phi_{ \alpha \mathbf{R}}  \rangle $, 
simply in terms of the generalization of the
transformation matrices $U_{\mathbf{k}}$ 
to possible non-unitarity matrices $M_{\mathbf{k}}$, that is
$M_{n \alpha  \mathbf{k} }
= \langle \psi_{ n \mathbf{k} }
\rvert \tilde{\psi}_{ \alpha \mathbf{k} } \rangle$, 
whereafter
  $\tilde{f}_{\alpha \beta \mathbf{k}} = \sum_n
M^{ \dagger }_{\alpha n \mathbf{k}} 
 f_{n \mathbf{k}} 
M_{n \beta \mathbf{k}}  $.
We use Latin and Greek letters to index 
orthonormal and nonorthogonal sets, respectively,
and implicitly sum over repeated index pairs.

In the nonorthogonal case, the 
density-matrix may be expanded in separable form via the tensor contraction
\begin{subequations}
\begin{align} \label{Eq:myrealngwf}
\hat{\rho} &{}= 
 \sum_{\mathbf{R}  \mathbf{R'} }
\lvert \phi_{\alpha \mathbf{R} }
 \rangle
 K^{ \alpha \beta}_{ 
\mathbf{R'} - \mathbf{R} }
\langle \phi_{\beta 
\mathbf{R'} } \rvert, \quad \mbox{where} \\
K^{ \alpha \beta}_{ \mathbf{R} } &{}=
 \frac{V_{\rm cell}}{\left( 
 2 \pi \right)^3 } 
  \int_{1\rm{BZ}} d\mathbf{k} \; e^{- i \mathbf{k} \cdot \mathbf{R} } 
 S^{ \alpha \gamma}
\tilde{f}_{\gamma \delta \mathbf{k}}
S^{ \delta \beta}, \label{Eq:myrealngwf2}
\end{align}
\end{subequations}
and the price to be paid for nonorthogonality is a
nontrivial metric tensor given by
$S_{\alpha \beta}
= \langle \phi_{\alpha \mathbf{R} } \rvert 
\phi_{\beta \mathbf{R'} } \rangle
\delta_{\mathbf{R} \mathbf{R'}}$, 
which defines the inter-relationship
between covariant vectors,
$\lvert \phi_{\alpha  \mathbf{R} } \rangle
= \lvert \phi^{\beta }_{ \mathbf{R} } \rangle 
S_{ \beta \alpha} $, 
and contravariant vectors (NGWF duals) 
$\lvert \phi^{ \alpha}_{\mathbf{R}} \rangle
= \lvert \phi_{\beta  \mathbf{R} } \rangle 
S^{ \beta \alpha} $.
The contravariant metric $S^{ \alpha \beta} $
in Eq.~\ref{Eq:myrealngwf2} is defined such that
$S_{ \alpha \gamma} 
S^{ \gamma \beta} 
 \equiv \delta_{\alpha}^{\;\; \beta}$, and is also independent  
 of the lattice vector.
Orthonormality is thus replaced by the
general tensor expression
 \begin{equation}  \langle \phi_{\alpha \mathbf{R} } \rvert 
\phi_{\gamma \mathbf{R'} } \rangle
S^{ \gamma \beta} =
S_{\alpha \gamma}
\langle \phi^{ \gamma}_{ \mathbf{R} } \rvert 
\phi^{\beta}_{ \mathbf{R'} } \rangle
= \delta_{\alpha}^{\;\; \beta} \delta_{\mathbf{R} \mathbf{R'}}.
\end{equation}

Numerous optimization procedures have been
developed for 
\emph{ab initio} Wannier functions. A widespread 
approach involves their construction in a post-processing
step, computing the $U_{n m \mathbf{k} }$ or 
$M_{n \alpha \mathbf{k} }$ matrices, and then
the generalized occupancies $\tilde{f}_{\mathbf{k}}$, 
\emph{following} the computation of 
the delocalized orbitals.
However, it has been recognized, and utilized in the context
of large-scale calculations for some time~\cite{PhysRevB.51.10157,
*0953-8984-14-11-303,*PhysRevB.47.9973,
*PhysRevB.50.4316,*PhysRevB.51.1456,
*PhysRevB.67.155108,*liu:1634}, that localized Wannier
functions may also be optimized directly \emph{in situ}, that is during
the process of solving for the electronic structure.
In the latter,
the basis expansion of the functions, 
$\left\lbrace \phi_{\alpha \mathbf{R}} 
\left( \mathbf{r} \right) \right\rbrace$ 
and the corresponding density
kernel $K_{\mathbf{R}}^{\alpha \beta}$ must be optimized
together, reconstructing the delocalized 
orbitals afterwards only if necessary.

A variety of plausible criteria may also be employed for
Wannier function optimization in either case, such as energy 
downfolding~\cite{PhysRevLett.96.166401} or
maximal Coulomb repulsion~\cite{PhysRevB.77.085122}, 
or, as used in this work, 
total-energy minimization or spatial localization.
Depending on their definition, 
these criteria may or may not uniquely 
define the Wannier functions, in that they may admit
some residual gauge freedom.
A particularly efficacious 
measure for localization is the second central moment
which, for a set of Wannier functions $\left\lbrace 
\lvert w_{n \mathbf{R} } \rangle \right\rbrace$ 
 takes the form of the  
\emph{spread} functional,
\begin{align}
\Omega &{}= \sum_n \left[ 
\langle  w_{ n \mathbf{0} }\rvert r^2 
\lvert  w_{ n \mathbf{0} }\rangle -
 \langle  w_{ n \mathbf{0} } \rvert \mathbf{r} \lvert 
 w_{ n \mathbf{0} } \rangle^2 \right] ,
\end{align}
where the generalization to the nonorthogonal case
does not yield straightforward physical interpretation.
MLWFs~\cite{PhysRevB.56.12847,*PhysRevB.65.035109} are 
those orthonormal Wannier functions generated
by unitary transformations 
$U_{n m \mathbf{k} } $ that minimize $\Omega$, for a fixed 
set of orbitals.
MLWFs are usually computed in a post-processing procedure, 
using an implementation such as \textsc{Wannier90}~\cite{wannier90},
and have been widely adopted as an accurate minimal basis 
with which to compute numerous ground-state and
excited-state properties, as well as to 
augment DFT with many-body 
interactions~\cite{wannierreview}.
MLWFs have been used to great effect, moreover, 
in the context of molecular dynamics, particularly interesting 
examples including the calculation of the dielectric permittivity and
dipolar correlation of liquid water~\cite{PhysRevLett.98.247401}, as
well as its dynamical charge and dipole tensors~\cite{PhysRevB.68.174302}.

\begin{figure}
\includegraphics[width=7cm]{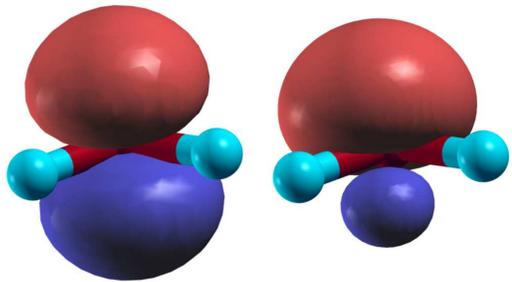}
\caption{(Color online) Wannier 
functions, NGWF (left) and MLWF (right), of 
predominantly oxygen 
$p_{z^2}$ (highest occupied, $1b_1$) character in 
H$_2$O, at zero applied field, 
with iso-surfaces at one sixth of their respective maxima.
Both types retain some residual arbitrariness following optimization.}
\label{Fig:WaterFunctions}
\end{figure}

NGWFs, unlike their orthonormal counterparts, are more commonly
optimized \emph{in situ}, as a by-product of 
total-energy minimization with respect to the density-matrix,
for example,
in the \textsc{ONETEP} linear-scaling DFT 
code~\cite{PhysRevB.66.035119,*onetep1,Hine20091041}.
In the latter, NGWFs are 
expanded in a fixed underlying
basis of periodic cardinal sine functions 
(also known as \emph{psinc}~\cite{mostofi-jcp03} or 
band-width limited $\delta$-functions),
whose spatial finesse is determined 
by a single variational parameter, 
the kinetic energy cutoff of 
the equivalent plane-wave basis.
The NGWFs are then those functions, 
when traced with their corresponding optimized density 
kernel, which reproduce the ground-state density-matrix, whence
the ground-state energy
\begin{align}
E_{0} &{}= \min_{n} E \left[ n \right] 
= \min_{\hat{\rho} } E \left[ \hat{\rho} 
\right]_{\hat{\rho} = \hat{\rho}^2 }  \\
&{}= \min_{ \mathbf{K}  , 
\{ \phi  \}  }
E \left[
 \mathbf{K}  , \{ \phi   \} 
\right]_{ \mathbf{K} = \mathbf{K}  \mathbf{S}  \mathbf{K}  }. \nonumber 
\end{align}

In practice, in order to 
extremize the total-energy with respect 
to idempotent density matrices,
two nested conjugate-gradients variational
minimization procedures are performed.
In the inner loop, the energy is minimized 
with respect to the elements of the density kernel,
for a fixed NGWF expansion, and in the outer, 
the density kernel is kept fixed while the 
total energy is minimized with respect 
to the NGWF psinc-expansion. 
A number of similar methods have been proposed
in which equations of motion generate optimized
nonorthogonal functions~\cite{PhysRevB.51.10157,
*0953-8984-14-11-303,*PhysRevB.47.9973,
*PhysRevB.50.4316,*PhysRevB.51.1456,
*PhysRevB.67.155108,*liu:1634}.

 An intuitive interpretation of Wannier functions 
is furnished via the modern theory of polarization~\cite{kingsmith, *resta},
in that changes in their centers 
$ \langle \mathbf{r} \rangle_{n m} = 
\langle  w_{n \mathbf{0} } 
\rvert \mathbf{r} \lvert  w_{m \mathbf{0} } \rangle$ 
exactly reproduce, and thus may be 
used to efficiently calculate, changes in the
polarization of insulating systems.
The change in electronic polarization 
$\delta \mathbf{P}$, subject to a gap-preserving perturbation,
 may be expressed as
 \begin{subequations}
\begin{align}
\delta \mathbf{P} {}&= 
- \frac{2 e}{V_{\rm cell}} \sum_n^{N} \delta 
 \langle \mathbf{r} \rangle_{n n} \;\;
\left( \mbox{if} \; \tilde{f}_{n m} = \delta_{n m}, n \le N \right),
  \\ 
 {}&=
- \frac{2 e}{V_{\rm cell}} 
\left[ \delta K_{\mathbf{0}}^{  \alpha \beta} 
 \langle  \mathbf{r} \rangle_{\beta \alpha} +
 K_{\mathbf{0}}^{ \alpha \beta} 
\delta \langle  \mathbf{r} \rangle_{\beta \alpha} \right] ,
\end{align}
\end{subequations}
where the $\mathbf{k}$-independence of the occupancies 
(also spin-degenerate) implies that  it is
sufficient to consider only the
$\mathbf{R} = \mathbf{0}$ term.
Here, respectively, we have provided the 
 orthonormal case for
$N$ occupied bands, 
and the more general, nonorthogonal case.

 It has been shown that close-to-orthonormal 
Wannier functions generated 
by means of direct minimization, of an appropriately
constructed functional, may be 
used to efficiently compute dielectric 
properties~\cite{PhysRevB.55.R1909,*PhysRevB.58.R7480}.
It is of importance, particularly for linear-scaling 
methods, to 
 generalize this result and verify that 
\emph{in situ} optimized NGWFs can reproduce
 electronic response properties 
with the same reliability as that of the well documented  MLWFs, as
NGWFs are increasingly being used in large-scale 
methods for  spectral partitioning
and dielectric properties, particularly in molecular 
systems~\cite{PhysRevB.84.165131,*PhysRevB.85.115404,
*PhysRevB.82.081102}.
The simplest such response property is perhaps
the high-frequency (termed ``clamped-ion" or ``static") linear dipole  
polarizability tensor 
\begin{equation}
\alpha_{i j} = \lim_{\omega \rightarrow \infty} 
\alpha_{i j} \left( \omega \right) =
\left. \frac{\partial P_i}{
\partial \mathcal{E}_j} \right|_{\delta \mathbf{R}_{\rm{ion}} = 0},
\end{equation}
where $\mathbf{\mathcal{E}}$ is an applied 
electric field within the dipole approximation.
This polarizability is somewhat different 
from that which is most frequently probed experimentally, 
namely the static or visual frequency regimes, and
neglects the response of the ionic positions.

\begin{table}
\begin{tabular}{l}
{\centering \begin{tabular}{|c|c|c|c|c|}
\hline\hline
$\bar{\alpha}$ 
&  NGWF & MLWF & Gaussian & Experiment \\
\hline
H$_2$O  & 10.58 &  10.47 & 10.76$^{a}$, 7.4$^b$ & 9.64$^{d}$, 9.79$^{f}$\\
NH$_3$  & 15.28 & 15.24 & 15.63$^{a}$, 12.1$^b$& 14.56$^{d}$, 18.9$^e$\\
CH$_4$   & 17.70 & 17.49 & 17.74$^{a}$, 14.8$^b$& 17.27$^{d}$, 17.5$^{e,f}$\\
C$_2$H$_4$ & 28.50 & 28.39 & 28.77$^{a}$, 25.6$^b$ & 27.70$^{d}$, 28.69$^{f}$\\
CO & 13.64 & 13.53 & 13.73$^{a}$, 12.1$^b$ & 13.09$^{d}$, 12.8$^e$, 13.16$^{f}$\\
CO$_2$ & 18.00 & 17.85 & 18.06$^{a}$, 14.8$^b$ & 17.51$^{d}$, 19.6$^{e,f}$, 17.48$^{f}$\\
N$_2$ & 11.99 & 11.87 & 12.31$^{a}$, 10.8$^b$ & 11.74$^{d,f}$, 11.5$^e$ \\
C$_{10}$H$_8$ & 122.8 & 123.0 &121.76$^{c}$ & 117.4$^{x,y}$, 118.9$^z$\\
\hline\hline
$ \kappa $ &  NGWF & MLWF & Gaussian & Experiment\\
\hline
H$_2$O  & 0.16 & 0.14 & 0.14$^{a}$ & 0.67$^{d}$\\
NH$_3$  & 2.45 & 2.64 & 2.70$^{a}$ & 1.94$^{d}$\\
C$_2$H$_4$ & 12.18 & 12.03 & 11.94$^{a}$ & 11.4$^{d}$\\
CO & 3.53 & 3.54 & 3.55$^{a}$ & 3.57$^{d}$\\
CO$_2$ & 13.88 & 13.96 & 13.70$^{a}$&  13.83$^{d}$, 13.70$^{f}$\\
N$_2$ & 4.50 & 4.55 & 4.83$^{a}$& 4.59$^{d}$, 4.45$^{f}$\\
C$_{10}$H$_8$ & 96.1 & 94.8 & 94.13$^{c}$ & 86.9$^{x}$, 79.0$^{y}$, 63.6$^z$\\
\hline\hline 
\end{tabular}\par} \\ 
\footnotesize{ $^{(a)}$ Static polarizability in a d-aug-cpVTZ 
basis~\cite{VanCaillie2000446}. } \\
\footnotesize{ $^{(b)}$ Static PBE polarizability in 6-311++G(d,p) basis at }\\ 
\footnotesize{ \phantom{$^{(b)}$} B3LYP/6-311G** optimised geometries~\cite{zope}. } \\
\footnotesize{ $^{(c)}$ Static polarizability in Sadlej pVTZ 
basis~\cite{hammond:144105}. } \\ 
\footnotesize{ $^{(d)}$ Compiled in Ref.~\onlinecite{VanCaillie2000446}, based on 
analysis of anisotropy data~\cite{Olney199759,*spackman:1288,*spackman2}.  } \\
\footnotesize{ $^{(e)}$ CRC Handbook~\cite{citeulike:1439705}.
$^{(f)}$ Extrapolated Rayleigh scattering~\cite{alms:3321}. } \\
\footnotesize{ $^{(x)}$ Anisotropic refraction~\cite{vuks}. 
$^{(y)}$ Laser Stark spectroscopy~\cite{heitz:976, vuks}. } \\ 
\footnotesize{ $^{(z)}$  Optical measurements at 
$632.8$~nm in solution~\cite{F29807601249}. }
\end{tabular}
\caption{Isotropic ($\bar{\alpha}$) and anisotropic ($\kappa$) 
polarizabilities (e$^2$~a$_0^2$~Ha$^{-1}$), 
from DFT using nonorthogonal 
(NGWF) and orthonormal (MLWF)
Wannier functions. 
Previous 
Gaussian-basis 
calculations~\cite{PhysRevLett.77.3865}, and 
experimental values are included.}
\label{Tab:Polarizabilities}
\end{table}

Two different Kohn-Sham DFT packages 
were used in order to compute polarizabilities
within the NGWF and MLWF formalisms, respectively
the \textsc{ONETEP}  linear-scaling code~\cite{PhysRevB.66.035119,*onetep1,Hine20091041},
and a combination of a plane-wave
pseudopotential package~\cite{QE}
and the \textsc{Wannier90}~\cite{wannier90} code.
An example of each type of function is depicted in Fig.\ref{Fig:WaterFunctions}.
A set of well-isolated, closed-shell molecules
were selected, so that a sawtooth-potential 
representation of the electric field could be used,
with the potential boundary maximally distant from the molecules,
up to a maximum field value of  
$\pm 8.0 \times 10^{-5}$~Ha~e$^{-1}$~a$_0^{-1}$,
in intervals of $2.0 \times 10^{-5}$~Ha~e$^{-1}$~a$_0^{-1}$,
for all systems.
The response remained well 
within the linear regime at these
field values, which lay well below the threshold for Zener
breakdown.
The rates of change in polarization 
was calculated using linear-regression of finite-difference data.
Identical norm-conserving 
pseudopotentials~\cite{opium}  were used in both cases, 
having been generated in the required formats, 
the Perdew-Burke-Ernzerhof (PBE) exchange-correlation 
approximation~\cite{PhysRevLett.77.3865}, 
and the same run-time parameters 
and analysis were used for both 
codes so far as possible.
Zero-field ground-state geometries were 
optimized using \textsc{ONETEP},
in cubic simulation cells of side length 
$40$~a$_0$ ($50$~a$_0$
in the case of naphthalene C$_{10}$H$_8$).
The density and potential were fully reset to those of 
atomic superpositions
upon each incrementation of the field.
An equivalent plane-wave cutoff of $1000$~eV, 
$\Gamma$-point Brillouin zone sampling, no
density-kernel truncation and NGWFs 
with a $10$~a$_0$ radius cutoff were used. 

In the case of nonaxially symmetric molecules, random 
initial guesses for the MLWFs were 
regenerated at each incrementation of the electric field.
For axially symmetric molecules such as
CO, CO$_2$ and N$_2$, however,
the maximal localization condition 
does not uniquely define the MLWF centers
under rotations about the axis, as discussed
in Ref.~\onlinecite{andrinopoulos:154105}.
While the sum of centers, and hence the 
transverse response, should be well-defined,
in practice this unbroken symmetry 
results in excessively noisy linear-response data.
It was found, however, that re-initializing 
the MLWFs to $s$-orbitals at each field value,
with centers coinciding with a set of
zero-field MLWFs, proved sufficiently 
robust to obtain excellent  linear fitting.
No such measures were 
necessary in the case of ONETEP NGWFs,
due to an effective symmetry 
breaking introduced by the underlying 
real-space \emph{psinc} grid.

The isotropic and anisotropic parts of 
the polarizability tensor $\mathbf{\alpha}$
are defined, respectively, as
\begin{align}
\bar{\alpha} = 
 \frac{ 1 }{3} \rm{tr} \left[ \mathbf{\alpha} \right], \quad \kappa = 
\sqrt{ \frac{3}{2} \rm{tr} \left[ \mathbf{\alpha}^2 \right]  - 
\frac{1}{2} \left( \rm{tr} \left[ \mathbf{\alpha} \right] \right)^2 },
\label{Eqn:polariz}
\end{align}
our computed values of which 
using NGWFs and MLWFs
are shown 
in Table~\ref{Tab:Polarizabilities}.
The quadratic mean fractional discrepancy 
between the isotropic parts 
was $8.0\times10^{-3}$, while the discrepancy
was greater for the anisotropic parts, at $4.1\times10^{-2}$.
As judged by the arithmetic mean fractional discrepancies 
(given henceforth in parentheses),
the NGWFs tended to provide slightly larger 
isotropic parts (by $6.6\times10^{-3}$), 
and also anisotropic parts (by $1.3\times10^{-3}$), than the MLWFs.
 Perhaps serendipitously, the NGWF values lay
closer than the MLWF results, 
for the isotropic parts, in all cases,
to the previous DFT(PBE) calculations 
of Ref.~\onlinecite{VanCaillie2000446}, 
computed using a sophisticated time-dependent coupled-perturbed method 
with a  triple-$\zeta$ Gaussian basis set; 
the quadratic (arithmetic) mean discrepancies with respect to 
these previous results were
$1.5\times10^{-2}$ ($1.3\times10^{-2}$) and 
$2.2\times10^{-2}$ ($2.0\times10^{-2}$), respectively.
The trend was reversed for anisotropies.

Polarizabilities calculated using the related
Wannier function varieties
agree rather well in spite of the
the significant technical dissimilarities between the 
\emph{ab initio} packages generating them, and there are a number 
of possible origins for the small discrepancies observed.
First considering the NGWF and MLWF values, both based on 
the plane-wave formalism and using the 
same ionic geometry,  the NGWF method is 
the more approximate in that it spatially truncates the Wannier functions
and the kinetic-energy operator.
Moreover, these methods differ in their handling of pseudopotentials, 
and, substantially, in their energy-minimization algorithms.
With respect to the previous Gaussian-basis results of 
Ref.~\onlinecite{VanCaillie2000446}, the possible origins
for discrepancy are manifold, most notably, the 
ionic geometries employed differ and the 
latter method treats the core electrons explicitly.

 The probable errors (arising from the linear
 fit to the data)
 in the isotropic and anisotropic
polarizabilities, denoted $\Delta \bar{\alpha}$
and $\Delta \kappa $, respectively, and  given by
\begin{subequations}
\begin{align}
\Delta \bar{\alpha}
&{}= \sqrt{
\sum_{i j} \left( \frac{\partial \bar{\alpha}}{\partial \alpha_{i j}} 
\Delta \alpha_{ i j} 
\right)^2 } =
\frac{1}{3} \sqrt{ \sum_i \left( \Delta \alpha_{i i } \right)^2 }, 
 \\ 
 \Delta \kappa  &{}=  \kappa^{-1}
 \sqrt{ \sum_{i j} \left[ \left(  
 \frac{3 \alpha_{j i }}{2}  - \sum_k \alpha_{k k} \frac{  \delta_{j i}}{2}
  \right) \Delta \alpha_{i j } 
 \right]^2 },
 \label{Eqn:estimator}
\end{align} 
\end{subequations}
were computed using 
the unbiased variance estimate 
$\left( \Delta \alpha_{i j} \right)^2$ 
on each polarizability component $\alpha_{i j}$,
and are shown in Table~\ref{Tab:Polarizabilities2}. 
The noise in the data for NGWFs is somewhat more
system dependent, 
as the NGWF truncation depends on the ionic geometry, 
and higher than in the MLWF case for most of the
molecules studied.
The quadratic (arithmetic) mean 
of the ratio of the estimated error
in the isotropic polarizability $\bar{\alpha}$
to its value was estimated at 
$1\times10^{-3}$ ($4\times10^{-4}$) for NGWFs and 
$2\times10^{-4}$ ($2\times10^{-4}$) for MLWFs.
Correspondingly, for the anisotropic part $\Delta \alpha $, we estimated
these ratios to be, 
respectively, $1\times10^{-2}$ ($7\times10^{-3}$) and
$3\times10^{-3}$ ($2\times10^{-3}$).
Nonetheless, the probable errors in the linear fits to the polarizability
data were extremely small using both methods,
and inconsequential with respect 
to the expected errors in  the 
approximate functional.

\begin{table}[b]
{\centering \begin{tabular}{|c|c|c|c|c|c|}
\hline\hline
$\Delta \bar{\alpha} / \bar{\alpha}$ 
&  NGWF  & MLWF & $\Delta \kappa / \kappa$ & NGWF  & MLWF  \\
\hline
H$_2$O  & $1 \times 10^{-4}$ & $8 \times 10^{-5}$ & NH$_3$  & $8 \times 10^{-3}$ & $8 \times 10^{-3}$ \\
NH$_3$  & $3 \times 10^{-3}$ & $5 \times 10^{-5}$ & NH$_3$  & $4 \times 10^{-2}$ & $5 \times 10^{-4}$ \\
CH$_4$   & $1 \times 10^{-4}$  & $ 5 \times 10^{-5}$  &  CH$_4$   &  -  & - \\
C$_2$H$_4$ & $3 \times 10^{-5}$  & $ 3 \times 10^{-5}$ & C$_2$H$_4$ & $ 2 \times 10^{-4}$ & $ 2 \times 10^{-4}$ \\
CO & $2 \times 10^{-4}$ &$3 \times 10^{-4}$ & CO & $2 \times 10^{-3}$ & $2 \times 10^{-3}$ \\
CO$_2$ & $2 \times 10^{-5}$  & $2 \times 10^{-4}$  & CO$_2$ & $5 \times 10^{-5}$  & $4 \times 10^{-4}$  \\
N$_2$ & $6 \times 10^{-6}$  & $3 \times 10^{-4}$  & N$_2$ &$5 \times 10^{-5}$  & $2 \times 10^{-3}$  \\
C$_{10}$H$_8$ & $4 \times 10^{-4}$ & $4 \times 10^{-4}$ & C$_{10}$H$_8$ & $1 \times 10^{-3}$ & $1 \times 10^{-3}$ \\
\hline\hline 
\end{tabular}\par} 
\caption{Probable fractional errors
in molecular electric polarizabilities computed using NGWFs and MLWFs.}
\label{Tab:Polarizabilities2}
\end{table}

In conclusion, we have shown that nonorthogonal
Wannier functions optimized \emph{in situ} may be
used to compute molecular polarizabilities
with an accuracy comparable to MLWFs post-processed
from plane-wave DFT. This result is promising
for the computation of numerous 
dielectric properties, and the full application of 
linear-scaling Wannier function analysis to 
large systems.  A promising avenue for future work
is the generalization of a method for the dielectric 
response in extended systems, such as that described
in Ref.~\onlinecite{PhysRevLett.73.712} and applied to solids in
Ref.~\onlinecite{PhysRevB.55.R1909,*PhysRevB.58.R7480}, 
to the linear-scaling NGWF formalism.

We are grateful to John Biggins and Danny Cole
for helpful discussions, and to
Mark Robinson and Peter Haynes for provision of software. 
D.D.O'R acknowledges the support of EPSRC 
and the National University of Ireland.  
M.C.P. acknowledges EPSRC support 
(Grants No. EP/G055904/1 and No. EP/F032773/1). 
A.A.M. acknowledges the support of 
EPSRC (EP/G05567X/1) and RCUK.

\end{document}